\documentclass[12pt]{article}
\usepackage{graphics} 
\def\lsim{\mathrel{\lower2.5pt\vbox{\lineskip=0pt\baselineskip=0pt
          \hbox{$<$}\hbox{$\sim$}}}}
\def\gsim{\mathrel{\lower2.5pt\vbox{\lineskip=0pt\baselineskip=0pt
          \hbox{$>$}\hbox{$\sim$}}}}
\def\real{\mathrel{\lower.0pt \hbox{$I\!\!R$}}}

\begin{document}   
\markright{Spacetime geometry of static fluid spheres\hfil}
\title{\Large \bf Spacetime geometry of static fluid spheres}
\author{Shahinur Rahman and Matt Visser\\[2mm]
{\small \it 
Physics Department, Washington University, 
Saint~Louis, Missouri 63130-4899, USA.}}
\date{{\small 16 March 2001; revised 28 January 2002; \LaTeX-ed \today}}
\maketitle
\begin{abstract}

We exhibit a simple and {\emph{explicit}} formula for the metric of an
arbitrary static spherically symmetric perfect fluid spacetime. This
class of metrics depends on one freely specifiable monotone
non-increasing generating function. We also investigate various
regularity conditions, and the constraints they impose. Because we
never make any assumptions as to the nature (or even the existence) of
an equation of state, this technique is useful in situations where the
equation of state is for whatever reason uncertain or unknown.

To illustrate the power of the method we exhibit a new form of the
``Goldman--I'' exact solution. This is a three-parameter closed-form
exact solution given in terms of algebraic combinations of
quadratics. It interpolates between (and thereby unifies) at least six
other reasonably well-known exact solutions.

\vspace*{5mm}
\end{abstract}
\vfill
\hrule
\bigskip
\centerline{\underline{E-mail:} {\sf srahman@artsci.wustl.edu}}
\centerline{\underline{E-mail:} {\sf visser@kiwi.wustl.edu}}
\bigskip
\centerline{\underline{Homepage:} 
{\sf http://www.physics.wustl.edu/\~{}visser}}
\bigskip
\centerline{\underline{Archive:} {\sf   gr-qc/0103065}}
\bigskip
\hrule
\clearpage
\def\Box{\nabla^2}
\def\d{{\mathrm d}}
\def\ie{{\em i.e.\/}}
\def\eg{{\em e.g.\/}}
\def\etc{{\em etc.\/}}
\def\etal{{\em et al.\/}}
\def\S{{\mathcal S}}
\def\I{{\mathcal I}}
\def\L{{\mathcal L}}
\def\R{{\mathcal R}}
\def\eff{{\mathrm{effective}}}
\def\Newton{{\mathrm{Newton}}}
\def\bulk{{\mathrm{bulk}}}
\def\matter{{\mathrm{matter}}}
\def\tr{{\mathrm{tr}}}
\def\normal{{\mathrm{normal}}}
\def\implies{\Rightarrow}
\def\half{{1\over2}}
\def\surface{{\mathrm{surface}}}
\section{Introduction}
\label{S:introduction}

The apparently simple problem of the general relativistic static
perfect fluid sphere has by now generated hundreds of scientific
articles. Good summaries of known results, and commentaries regarding
the extant literature can be found in the book by Kramer
{\etal}~\cite{Exact}, and in the recent review articles by Delgaty and
Lake~\cite{lake}, and Finch and Skea~\cite{skea}.

One of the more common approaches (certainly not the only approach) is
to pick some barotropic equation of state $\rho=\rho(p)$, pick the
central pressure, apply the Tolman--Oppenheimer--Volkoff equation, and
integrate outwards until one reaches the surface of the ``star''
(assumed to be characterized by the innermost zero-pressure surface
$p=0$).  Now there are many physical situations in which one simply
does not know the equation of state, either because of uncertainties
in the basic physics (for example, there are still some uncertainties
regarding the equation of state for nuclear matter in neutron stars),
or more prosaically because the chemical composition of the ``star''
may vary throughout its bulk so that it is not meaningful to speak of
a single equation of state for the entire body.\footnote{%
A somewhat different approach, explored by Baumgarte and
Rendall~\cite{Baumgarte}, consists of specifying a non-negative but
otherwise arbitrary density profile and then integrating the
Tolman--Oppenheimer--Volkoff equation (without assuming any equation
of state) to determine the pressure profile. Such a procedure
generates geometries more general than the more standard approach
sketched above.}

We therefore decided to see what explicit constraints on the spacetime
geometry could be deduced directly from the perfect fluid condition,
without reference to any particular equation of state. To start with,
note that by using the coordinate freedom inherent in general
relativity any static spherically symmetric geometry can be put into a
form where there are only two independent metric components, typically
functions of the radial coordinate. The most common such forms are
given by Schwarzschild coordinates (area coordinates, curvature
coordinates)
\begin{equation}
\d s^2 = 
- |\tilde g_{tt}(\tilde r)| \;\d t^2 + 
   \tilde g_{rr}(\tilde r) \;\d \tilde r^2 + 
\tilde r^2 (\d\theta^2 + \sin^2\theta \; \d\phi^2),
\end{equation}
and isotropic coordinates
\begin{equation}
\label{E:basic-isotropic}
\d s^2 = 
- |g_{tt}(r)| \;\d t^2 
+ g_{rr}(r) \left[ \d r^2 + r^2 (\d\theta^2 + \sin^2\theta \; \d\phi^2) \right].
\end{equation}
Now spherical symmetry by itself automatically implies that once one
calculates the Einstein tensor and goes to an orthonormal
frame\footnote{
Hatted indices will always be used to denote the orthonormal frame
attached to a particular coordinate system.}
%
\begin{equation}
G_{\hat\theta\hat\theta} = G_{\hat\phi\hat\phi}.
\end{equation}
If the geometry is to represent a perfect fluid then in addition we
demand pressure isotropy
\begin{equation}
G_{\hat r\hat r} = G_{\hat\theta\hat\theta} = G_{\hat\phi\hat\phi}.
\end{equation}
This places a single differential constraint on the metric components,
and so we expect the class of metrics representing a perfect fluid
geometry to have only one freely specifiable metric component --- more
precisely, we expect there to be a single freely specifiable
generating function, call it $z(r)$, that should characterize the
entire class of metrics
\begin{equation}
g[z(r)]
\end{equation}
for static perfect fluid spheres. Since the pressure isotropy
condition involves derivatives of the metric components, we expect the
metric $g[z(r)]$ to be some {\emph{functional}} of the generating
function $z(r)$, unavoidably involving derivatives and
integrations. These comments are of course quite standard and in some
form or another implicitly underlie all extant static spherically
symmetric perfect fluid solutions. The novelty in the current article
lies in the fact that we will make this implicit procedure explicit
and thereby will be able to exhibit the most general form of the
metric for static spherically symmetric perfect fluid spacetimes.
That is, we are seeking an explicit closed-form
(algebraic-integro-differential) solution to the pressure isotropy
condition.

We report that an explicit and relatively simple characterization of
this type does in fact exist. It involves a single derivative, some
algebraic manipulations (of which the worst is taking a square root)
and an explicit integration. The technique can be viewed as a simple
algorithm for constructing {\emph{all}} static spherically symmetric
perfect fluid geometries.\footnote{%
In particular our technique provides a way of algorithmically
generating all possible Baumgarte---Rendall
configurations.~\cite{Baumgarte}}
We also discuss the restrictions that must
be placed on the generating function in order to get ``physically
reasonable'' geometries. 

Finally we present a few specific examples where we demonstrate how
various well-known solutions fit into our scheme. We exhibit a
particularly striking three-parameter perfect-fluid solution, given in
closed form in terms of algebraic functions.  The solution is
presented in a new manner, and is with hindsight equivalent to the
Goldman-I solution (Gold-I solution in the Delgaty--Lake
classification), which we show is in turn equivalent to the
Glass--Goldman solution (G-G solution). Furthermore, in various
regions of parameter space the general solution reduces to at least
six different previously derived solutions. In particular our solution
includes three two-parameter sub-solutions: the interior Schwarzschild
solution, the Stewart solution, and (in the Delgaty--Lake
classification) the Kuch5 XIII solution. It also contains, as
one-parameter branches, the Einstein, de Sitter, and anti-de Sitter
solutions. We do not claim this list is exhaustive.

\section{Perfect fluid spheres}
\setcounter{equation}{0}

Consider a spherically symmetric static spacetime geometry. Without
loss of generality we know we can put it into isotropic coordinates
\begin{equation}
\d s^2 = 
- |g_{tt}(r)|  \; \d t^2 
+ g_{rr}(r) 
\left[ \d r^2 + r^2 (\d\theta^2 + \sin^2\theta \; \d\phi^2) \right].
\end{equation}
Our first key result can be phrased as a simple theorem:

\subsection*{Theorem I}

Pick an \emph{arbitrary} non-increasing function $z(r)$ [that is: a
suitably smooth function $z(r)$ with $z'(r) \leq 0$], introduce a
dummy integration variable $\bar r$, and formally construct the metric
\begin{eqnarray}
\d s^2 &=& 
- \exp\left\{ 
\pm 2 \int^r {\sqrt{- \bar r z'}\over1-z \bar r^2} \;\d\bar r 
\right\} \d t^2
\nonumber
\\
&& \quad
+ \exp\left\{ 
-2 \int^r {\pm\sqrt{-\bar r z'}-2 \bar r z\over1-z \bar r^2} \; \d\bar r 
\right\} \;
\left[ \d r^2 + r^2 (\d\theta^2 + \sin^2\theta \; \d\phi^2) \right].
\label{E:master}
\end{eqnarray}
Then this metric is guaranteed to be real [by the non-increasing
property of $z(r)$] and always describes a static spherically
symmetric distribution of perfect fluid matter. Conversely, the
spacetime metric generated by any static spherically symmetric
distribution of perfect fluid matter can be put into this form for
some suitable non-increasing $z(r)$.

\subsubsection*{Proof $\Rightarrow$}

By explicit computation
\begin{equation}
G_{\hat r\hat r} = G_{\hat\theta\hat\theta} = G_{\hat\phi\hat\phi} =
{ (z r^4 )' \over g_{rr} \;(1 - z r^2)^2 \; r^3}.
\label{E:geometric}
\end{equation}
The computations have been carried out and cross checked using a
combination of pencil and paper, the CARTAN~\cite{cartan} package
under Mathematica,\footnote{\sf http://www.wolfram.com} and the
standard distribution of Maple.\footnote{\sf http://www.maplesoft.com}
Invoking the Einstein equations this purely geometric statement
(\ref{E:geometric}) implies
\begin{equation}
p = {1\over8\pi G_{\Newton}}\;
{ (z r^4 )' \over g_{rr} \; (1 - z r^2)^2 \; r^3}.
\label{E:pressure}
\end{equation}
QED.

\subsubsection*{Proof $\Leftarrow$}

Suppose, on the other hand, we start with a static spherically
symmetric perfect fluid. Without loss of generality we can put the
metric in isotropic coordinates and choose the coefficients to be
\begin{equation}
\d s^2 = 
- \exp\{-2\varphi(r)\} \; \d t^2 
+ \exp\{+2\varphi(r)+4\psi(r)\} \;
 \left[ \d r^2 + r^2 (\d\theta^2 + \sin^2\theta \; \d\phi^2) \right].
\end{equation}
Then
\begin{eqnarray}
G_{\hat t\hat t} &=&  -{1\over g_{rr}} \left[
2\varphi'' + 4 \psi'' + (\varphi')^2 + 4 (\psi')^2 + 4 \psi' \varphi'  + 
{4 \varphi'\over r} + {8 \psi'\over r}
\right];
\\
G_{\hat r\hat r} &=& \hphantom{+}{1\over g_{rr}} 
\left[ 4(\psi')^2 + {4\psi'\over r} - (\varphi')^2 \right];
\\
G_{\hat\theta\hat\theta} &=& G_{\hat\phi\hat\phi} = 
{1\over g_{rr}} \left[ 2\psi'' + {2\psi'\over r} + (\varphi')^2 \right].
\end{eqnarray}
Demanding pressure isotropy yields the equation
\begin{equation}
 (\varphi')^2 + \psi'' - 2(\psi')^2 - {\psi'\over r} = 0,
\end{equation}
which is easily solved {\emph{algebraically}} (for the {\em
derivative} $\varphi'$)
\begin{equation}
\varphi' = \pm \sqrt{  2(\psi')^2 + (\psi')/r - \psi'' }.
\label{E:isotropy1}
\end{equation}
We could satisfy this equation by picking a ``generating function''
$\Theta(r)$ and setting
\begin{eqnarray}
\psi'(r) &=& \Theta(r);
\\
\varphi'(r) &=& \pm  \sqrt{  2\Theta(r)^2 + \Theta(r)/r - \Theta'(r) }.
\end{eqnarray}
But with this particular choice of generating function it is difficult
to guarantee the reality of the resulting metric. Instead we find it
more useful to make the algebraic definition
\begin{equation}
\psi'(r) = {z(r) r \over 1 - z(r) r^2 }; 
\qquad
\hbox{that is}
\qquad
z(r) = {\psi'(r)\over r[1+r\psi'(r)]}.
\end{equation}
With this definition for the generating function $z(r)$ it is now a
simple matter to verify that the isotropy condition
(\ref{E:isotropy1}) is equivalent to
\begin{equation}
\varphi'(r) = \pm {\sqrt{-r z'}\over1-z r^2}.
\end{equation}
Integrating and substituting, we get the form of the metric given in
the statement of the theorem, with now a very simple condition on
$z(r)$ [the non-increasing condition] being sufficient to guarantee
reality of the metric.
\\*
QED.

\subsubsection*{Aside:}

We also mention, because it is relatively simple, that for this entire
class of metrics
\begin{equation}
p' = \mp  {\sqrt{-r z'}\over 1- z r^2} \; (\rho + p).
\end{equation}
The choice of sign for the square root will be fixed later on, once we
demand positivity of density at the origin.  In contrast, we note that
the corresponding formula for $G_{\hat t\hat t}$ is quite messy. It is
better, but still less than ideal, to consider $G_{\hat t\hat t} + 3
\, G_{\hat r\hat r} = 2 R_{\hat t\hat t}$ which can be cast into any
of the equivalent forms
\begin{eqnarray}
G_{\hat t\hat t} + 3 \, G_{\hat r\hat r}  
&=&
\mp{5 z' + 3 r^2 z z' + 2 r^3 (z')^2 + r (1 -z r^2) z''
\over g_{rr} \;\sqrt{-r z'} \; (1- z r^2)^2}
\label{E:primary}
\\
&=&
\mp{4 z' +  (1 -z r^2)^3 [r  (1 -z r^2)^{-2} z']' 
\over g_{rr} \; \sqrt{-r z'} \; (1- z r^2)^2}
\\
&=&
\pm{2\over g_{rr} \; (1-z r^2) \; r^2} 
\left[ (r^2 \sqrt{-r z'})' +  \sqrt{-r z'} {(z r^4)'\over 1- z r^2} \right]
\\
&=&
\pm{2\over g_{rr}} 
\left[
{(\sqrt{-r z'})'\over 1- z r^2} +  
{2\sqrt{-r z'} \over r (1- z r^2)^2}  +
\sqrt{-r z'} \left({1\over 1- z r^2}\right)' \right]
\\
&=&
\pm{2\over g_{rr}} 
\left[
\left({\sqrt{-r z'}\over 1- z r^2}\right)' +  
{2\sqrt{-r z'} \over r (1- z r^2)^2}
\right]
\\
&=&
\pm{2\over g_{rr}} 
\left[
{1\over r^2}\left({r^2\sqrt{-r z'}\over 1- z r^2}\right)' +  
{2z r \sqrt{-r z'} \over (1- z r^2)^2}
\right].
\end{eqnarray}
We agree that none of these formulae are stunningly pleasant, but
despite considerable effort this is (in general) the best we have been
able to do.

Once we apply the Einstein equations
\begin{equation}
\rho + 3 p =  \pm {2\over8\pi G_{\Newton}\; g_{rr} } 
\left[
{1\over r^2}\left({r^2\sqrt{-r z'}\over 1- z r^2}\right)' +  
{2z r \sqrt{-r z'} \over (1- z r^2)^2}
\right].
\end{equation}

\subsubsection*{Comment:}

We have not used any equation of state anywhere in the
derivation. Furthermore, we have not yet applied any regularity
conditions to the metric --- so far it could represent a perfect fluid
sphere such as a star, or a completely liquid planet (\eg, Jupiter?).
It could also represent the fluid portion of a mostly liquid planet
surrounding a solid core (\eg, Saturn), or a black hole surrounded by
a spherically symmetric perfect fluid halo (an example of a so-called
``dirty black hole''~\cite{dbh} or ``hairy black hole''), or even a
traversable wormhole~\cite{morris,visser} supported by ``exotic''
perfect fluid.

We also wish to contrast this {\em explicit} formula for the metric
[equation (\ref{E:master})] with the more traditional {\em implicit}
formulation of the problem. (Typically along the lines of ``solve a
certain differential equation for one of the metric components and
implicitly substitute the result back into the metric ansatz''.)

\section{Regularity conditions}
\setcounter{equation}{0}

We now investigate the effect of placing various regularity conditions
on the geometry and the fluid.

\subsection{Regularity of the geometry at the origin}

If we focus on what is perhaps the astrophysically most interesting
case, that of a perfect fluid star (or a completely liquid planet),
then we want to impose some regularity conditions at the centre. At a
minimum we want the geometry to be regular, which at the most
elementary level requires~\cite{lake}
\begin{equation}
g_{tt}(r=0) = \hbox{finite}; \qquad g_{tt}'(r=0) = 0;
\end{equation}
and
\begin{equation}
g_{rr}(r=0) =  \hbox{finite}; \qquad g_{rr}'(r=0) = 0.
\end{equation}
We can then without loss of generality rescale $r$ to set $g_{rr}(r=0)
= 1$ (this only works because we are using isotropic coordinates); it
is convenient to {\emph{not}} rescale $g_{tt}(r=0)$. Then these
geometric regularity conditions can be satisfied by: (1) specifying the
lower limit of integration to be the origin; then (2) setting the
integration constants by defining
\begin{eqnarray}
\d s^2 &=& 
- \exp\{-2\phi(0)\}\;
  \exp\left\{ 
  \pm 2 \int_0^r {\sqrt{-\bar r z'}\over1-z \bar r^2} \d\bar r 
  \right\} \d t^2
\nonumber
\\
&& \quad
+ \exp\left\{
  -2 \int_0^r {\pm\sqrt{-\bar r z'}-2 \bar r z\over1-z \bar r^2} \d\bar r 
   \right\}  \;
\left[ \d r^2 + r^2 (\d\theta^2 + \sin^2\theta \; \d\phi^2) \right];
\label{E:regularity}
\end{eqnarray}
and finally (3) demanding that both
\begin{equation}
{\sqrt{-r z'}\over1-z r^2} \to 0 
\qquad
\hbox{and}
\qquad
{ r z\over1-z r^2} \to 0
\qquad
\hbox{as}
\qquad
r \to 0.
\end{equation}
This requires both $z(r)$ and $z'(r)$ to be finite at the origin.

\subsection{Finiteness of central pressure and density}

The central pressure, derived from equation (\ref{E:pressure}) using
the condition of geometric regularity at the origin, is
\begin{equation}
p_c = {4 \; z(0)\over8\pi G_\Newton},
\end{equation}
which gives no additional constraint beyond regularity of the geometry
itself. Indeed
\begin{equation}
z(0) = 2 \pi  G_\Newton \; p_c.
\end{equation}
  
On the other hand, by considering $\rho+3p$ as one nears the origin we
can derive additional constraints. Suppose we expand $z(r)$ in a power
series
\begin{equation}
z(r) = z(0) + z'(0) r + \half z''(0) r^2 + O(r^3).
\end{equation}
Then, evaluating the numerator and denominator of equation
(\ref{E:primary}) separately
\begin{equation}
[\rho+3p](r) =  {\mp1\over8\pi G_\Newton} 
{5 z'(0) + 6 z''(0) r + O(r^2) 
\over
\sqrt{-r z'(0) - z''(0) r^2} ( 1+ O(r))}.
\end{equation}
So if the central value of  $\rho+3p$ is to be finite we must have
\begin{equation}
z'(0) = 0,
\end{equation}
in which case 
\begin{equation}
[\rho+3p](r) =  {\mp1\over8\pi G_\Newton} 
{6 z''(0) + O(r) 
\over
\sqrt{- z''(0)}( 1+ O(r))} 
=
{\pm1\over8\pi G_\Newton} 
6 \sqrt{- z''(0)}  + O(r).
\end{equation}
Thus if both pressure and central density are to be {\em finite} we must have
\begin{equation}
z(r) \approx z(0) + \half z''(0) r^2 + O(r^3)
\end{equation}
with
\begin{equation}
\rho_c+3p_c = {\pm6\over8\pi G_\Newton} \sqrt{{-z''(0)}}.
\end{equation}
So in summary, finiteness of central pressure and density implies
\begin{equation}
z(0) = {\mathrm{finite}},
\end{equation}
while
\begin{equation}
z'(0) = 0,
\end{equation}
and
\begin{equation}
z''(0) = - {(8\pi G_\Newton)^2\over 36} [\rho_c+3p_c]^2.
\end{equation}

\subsection{Positivity of central pressure and density}

For the central pressure to be {\em positive} we additionally  require
\begin{equation}
z(0) > 0.
\end{equation}

For the central density to be positive we need, first, to take the
positive root in the expression for $\rho+3p$. This implies a specific
choice for the $\pm$ throughout the entire body of the ``star''. That is,
we must take
\begin{eqnarray}
\label{E:positive-root}
\d s^2 &=& 
-  \exp\{-2\phi(0)\}\;
   \exp\left\{ 
   2 \int_0^r {\sqrt{-\bar r z'}\over1-z \bar r^2} \d\bar r 
   \right\} \d t^2
\nonumber
\\
&& \quad
+ \exp\left\{ 
  -2 \int_0^r {\sqrt{-\bar r z'}-2 \bar r z\over1-z \bar r^2} \d\bar r 
  \right\} \;
\left[ \d r^2 + r^2 (\d\theta^2 + \sin^2\theta \; \d\phi^2) \right].
\end{eqnarray}
Second, we must demand
\begin{equation}
\sqrt{-z''(0)} > 2 z(0).
\end{equation}
That is
\begin{equation}
z''(0) < - 4 z(0)^2.
\end{equation}

\subsection{Positivity of pressure (and density?)}

Enforcing positivity of pressure is easy: the pressure is proportional to $(z
r^4)'$ multiplied by quantities that are guaranteed positive. So the
pressure will be positive as long as
\begin{equation}
(z r^4)' > 0.
\end{equation}
The innermost surface at which $(z r^4)' = 0$ is defined to be the
surface of the star, denoted by $r_\mathrm{surface}$.

The positivity of pressure throughout the star then implies
\begin{equation}
z(r) + 4 {z'(r)\over r} > 0;
\qquad \implies \qquad
z(r) > - 4 {z'(r)\over r} > 0.
\end{equation}
Thus the function $z(r)$ must be positive at least as far out as the
surface of the star.

Guaranteeing positivity of density throughout the star is much more
difficult to achieve: Mathematically this is because the density (in
contrast to the pressure) depends on second derivatives $z''(r)$ of
the generating function $z(r)$. Physically this arises because we have
not specified any equation of state. Because of this it should not be
too surprising that we cannot say everything about the ``star'' ---
the surprise perhaps is just how much we can do without an equation of
state.

\subsection{Monotonic decrease of pressure and density?}

Similarly, guaranteeing a monotone decrease of pressure and density
throughout the star is generically difficult to achieve.  On the other
hand, what can be done very easily is to derive conditions for the
gravitational potential to be monotone increasing as we move from the
center (that is, to keep the local gravitational force pointing
downwards). Given the sign choice made to keep the central density
positive, we need now only add the condition
\begin{equation}
z(r) \; r^2 < 1
\end{equation}
throughout the interior of the star. (This condition also prevents
singularities in the integration used to define the metric).

But given our sign choice for the root, we already know
\begin{equation}
p' = -  {\sqrt{-r z'}\over 1- z r^2} \; (\rho + p).
\end{equation}
Thus, assuming monotonicity for the gravitational potential implies
\begin{equation}
\hbox{sign}(p') = - \hbox{sign}(\rho+p).
\end{equation}
So under these assumptions the pressure will be monotone decreasing if
and only if the null energy condition (NEC) is
satisfied~\cite{nec-sec}.

\subsection{Positivity of total mass}

To calculate the total mass out to radius $r$ we use the
Hernandez--Misner mass formula~\cite{Misner}
\begin{equation}
m(r) 
= {1\over2} R_{\hat\theta\hat\phi\hat\theta\hat\phi}\; \; g_{\theta\theta}{}^3
= {1\over2} 
{R_{\theta\phi\theta\phi}\over g_{\theta\theta}\; g_{\phi\phi} } \; \;
g_{\theta\theta}{}^3.
\end{equation}
This formula is valid for any spherically symmetric spacetime in any
spherically symmetric coordinate system~\cite{Misner}. Applied to the
metric (\ref{E:positive-root}) in isotropic coordinates one obtains
\begin{equation}
m(r) 
=
\sqrt{g_{\theta\theta}(r)} \; r\; 
\left\{
{2\sqrt{-r \, z'}(1+z(r)\,r^2) - 4 r\,z(r) + z'(r)\, r^2
\over
2(1-z(r)\,r^2)^2}
\right\}.
\end{equation}
The surface of the star is located at $r_\surface$, where $(z r^4)' =
0$. In particular $z'(r_\surface)=-4 z_\surface/r_\surface$. Then a
brief computation shows that the total mass of the ``star'' is
\begin{equation}
m(r_\surface) 
=
\sqrt{g_{\theta\theta}(r_\surface)} \; r_\surface \; 
\left\{
{2\sqrt{z_\surface}(1+z_\surface \,r_\surface^2) - 4 z_\surface \,r_\surface
\over
(1-z_\surface\,r_\surface^2)^2}
\right\}.
\end{equation}
Both numerator and denominator can be factorized to yield
\begin{equation}
m(r_\surface) 
= 
\sqrt{g_{\theta\theta}(r_\surface)} \; 
{2 \;r_\surface \; \sqrt{z_\surface}
\over
(1+\sqrt{z_\surface}\,r_\surface)^2} 
= R_\surface \; {2 \;r_\surface \; \sqrt{z_\surface}
\over
(1+\sqrt{z_\surface}\,r_\surface)^2}. 
\end{equation}
Here $R_\surface = \sqrt{g_{\theta\theta}(r_\surface)}$ is radius of
the ``star'' in curvature coordinates (Schwarzschild coordinates).
Since each of these factors is manifestly positive, so is the total
mass.

The ``compactness'' can be defined by
\begin{equation}
\chi \equiv {2\,m(r_\surface)\over R_\surface} 
= {4 \;r_\surface \; \sqrt{z_\surface}
\over
(1+\sqrt{z_\surface}\,r_\surface)^2}. 
\end{equation}
Since $4 x/(1+x)^2 \leq 1$, we have $\chi\in[0,1]$ and the very
sensible result
\begin{equation}
m(r_\surface) \leq R_\surface/2.
\end{equation}
That is:
\begin{equation}
R_{\mathrm{Schwarzschild}} = 2\, m(r_\surface) = 
{2 \;G_\Newton \; M_{\mathrm{physical}}}  
\leq R_\surface.
\end{equation}
These observations are not enough to guarantee that the density is
everywhere positive, but they do place powerful constraints on its
behaviour.

\subsection{Volume-averaged strong energy condition}

Another simple constraint on the stress-energy distribution for a
static perfect fluid sphere that can be extracted without specifying
an equation of state is a certain type of ``weighted volume average''
of the strong energy condition.  Specifically, consider
\begin{equation}
\int g_{rr} \left\{ \rho + 3p \right\} r^2 \d r =
{1\over4\pi G_{\Newton} } \int r^2 \d r 
\left[
{1\over r^2}\left({r^2\sqrt{-r z'}\over 1- z r^2}\right)' +  
{2z r \sqrt{-r z'} \over (1- z r^2)^2}
\right].
\end{equation}
We have already chosen the positive sign for the root, since we want
the central density to be positive. Note that this is not the ``proper
volume average'' (which would correspond to $\int g_{rr}{}^{3/2} \;r^2
\; \d r$) but is instead weighted by a compensating factor 
$g_{rr}{}^{-1/2}$, chosen to simplify the mathematical
analysis. (The occurrence and usefulness of this and related
{\emph{weighted}} volume averages is quite common in spherically
symmetric systems.)

Then, using the regularity conditions we have already deduced for the
origin, for any value of $r_*$ we deduce
\begin{equation}
\int_0^{r_*} g_{rr} \left\{ \rho + 3p \right\} r^2 \d r =
{1\over4\pi G_{\Newton} }
\left[
\left.\left({r^2\sqrt{-r z'}\over 1- z r^2}\right)\right|_{r_*} 
+
2 \int r^3 \d r 
{z \sqrt{-r z'} \over (1- z r^2)^2}
\right].
\end{equation}
The first term is non-negative by the non-increasing property of
$z(r)$, which was required just to enforce reality of the metric (plus
the constraint $z(r)r^2 < 1$, which was imposed to keep the local
gravitational force pointing downwards). The second term is positive
definite for the same reason, so we have, for all $r_*$
\begin{equation}
\int_0^{r_*} g_{rr} \left\{ \rho + 3p \right\} r^2 \d r > 0.
\end{equation}
This is not the strong energy condition (SEC, $\rho+3p>0$) itself, but
is at least a weighted volume average thereof~\cite{nec-sec}. This
implies that the energy density is not permitted to become too
violently negative.

\subsection{Subluminal speed of sound?}

It is traditional to compute the quantity 
\begin{equation}
\left.\left({\d p\over\d \rho}\right)\right|_{\mathrm{fluid}} 
\equiv
{\d p\over\d r} \left[{\d \rho\over\d r}\right]^{-1}
\end{equation}
and to demand that this be less than or equal to $c^2$, on the grounds
that this quantity is alleged to represent the physical speed of sound
(which certainly should be subluminal). This assertion is dangerously
misleading, and cannot be justified without significant additional
technical assumptions above and beyond those that have so far been
made.

Specifically, let us assume that the fluid is described by some
equation of state
\begin{equation}
p = p(\rho,X).
\end{equation}
Here $X$ stands for some collection of variables characterizing the
fluid, possibly chemical concentrations, entropy density, temperature,
or the like. Then
\begin{equation}
{\d p\over\d r} = 
\left.{\partial p\over\partial \rho}\right|_X {\d \rho\over\d r} +  
\left.{\partial p\over\partial X}\right|_\rho {\d X\over\d r}.
\end{equation}
Thus
\begin{equation}
\left.\left({\d p\over\d \rho}\right)\right|_{\mathrm{fluid}}  = 
\left.{\partial p\over\partial \rho}\right|_X +  
\left.{\partial p\over\partial X}\right|_\rho {\d X\over\d r} 
\left[{\d \rho\over\d r}\right]^{-1}.
\end{equation}
That is
\begin{equation}
\left.\left({\d p\over\d \rho}\right)\right|_{\mathrm{fluid}}  = 
c^2_s(X) +  
\left.{\partial p\over\partial X}\right|_\rho
\left.\left({\d X\over\d \rho}\right)\right|_{\mathrm{fluid}}.
\end{equation}
In other words $(\d p/\d\rho)_{\mathrm{fluid}}$ can be related to the
(constant $X$) speed of sound $c_s(X)$ {\emph{if and only if}} you add
extremely powerful additional assumptions. (Such as $\partial
p/\partial X =0$, implying an a priori exactly barotropic equation of
state. Or $\d X/\d r = 0$, implying for instance either thorough
mixing of the entire fluid mass, an adiabatic star, or an isothermal
star.)  Without such additional assumptions no particular conclusion
regarding the relationship between $(\d p/\d\rho)_{\mathrm{fluid}}$
and the physical speed of sound can be drawn~\cite{mtw}.  Given our
philosophy in this article (we wish to see what can be deduced without
making assumptions about the equation of state) such assumptions would
be completely opposite to our purpose, and so we do not seek to impose
the condition $\d p/\d\rho \leq c^2$.

\subsection{Summary}

We can summarize the essential core of these regularity conditions in
the following theorem:

\subsection*{Theorem II}

Let $z(r)$ be a positive non-increasing function ($z'(r)\leq0$) such
that:
\begin{enumerate}
\item $z(0)$ is finite;
\item $z'(0)=0$;
\item $z''(0) < - 4 z(0)^2$;
\item $z(r) < 1/r^2$;
\end{enumerate}
and consider the metric (guaranteed to be real)
\begin{eqnarray}
\d s^2 &=& 
-  \exp\{-2\phi(0)\}\;
   \exp\left\{ 
   2 \int_0^r {\sqrt{-\bar r z'}\over1-z \bar r^2} \d\bar r 
   \right\} \d t^2
\nonumber
\\
&& \quad
+ \exp\left\{ 
  -2 \int_0^r {\sqrt{-\bar r z'}-2 \bar r z\over1-z \bar r^2} \d\bar r 
  \right\} \;
\left[ \d r^2 + r^2 (\d\theta^2 + \sin^2\theta \; \d\phi^2) \right].
\label{E:theorem2}
\end{eqnarray}
Then this metric represents a static perfect fluid sphere with:
\begin{enumerate}
\item regular geometry at the origin;
\item finite and positive pressure and density at the origin;
\item a local gravitational field that always points downward.
\end{enumerate}
Conversely, any static perfect fluid sphere satisfying these last three
conditions can be cast into the preceding form with a generating
function $z(r)$ satisfying the first four conditions.

\bigskip

Furthermore, under the conditions enunciated above, if the system is
additionally isolated (so that the pressure drops to zero at some
finite radius), then the total mass is guaranteed to be both positive
and bounded.

\subsection*{Proof $\Leftrightarrow$}

This theorem is just a codification of the most salient of the
preceding results.
\\* QED.

\section{Examples:}
\setcounter{equation}{0}

While the metric given in equation (\ref{E:theorem2}) is guaranteed to
be a perfect fluid for a very wide class of generating functions
$z(r)$, it is only for a much more restricted class of generating
functions that the relevant integrals can be performed in terms of
elementary functions. We now present several examples where this can
be done.

\subsection{Schwarzschild Exterior Geometry}
\label{S:exterior}

The Schwarzschild exterior solution corresponds to
\begin{equation} 
z(r) = {(m/2)^2\over r^4}; \qquad m = G_\Newton \; M_{\mathrm{physical}},
\end{equation}
together with choosing the positive sign for the root. A brief
computation leads to $\rho=0$, $p=0$, and the isotropic form of the
Schwarzschild exterior metric.
\begin{equation}
\d s^2 = 
- 
\left({1-m/(2r)\over1+m/(2r)}\right)^2 \d t^2 
+
\left(1+{m\over2r}\right)^4 
\left[ \d r^2 + r^2 (\d\theta^2 + \sin^2\theta \; \d\phi^2 )\right].
\end{equation}
Note that the exterior Schwarzschild does {\emph{not}} satisfy the
regularity requirements for a ``normal'' fluid sphere --- in
particular $r=0$ is not a ``point'' but instead corresponds (in these
isotropic coordinates) to a second asymptotically flat region. Because
the geometry is not regular at the origin we cannot use equation
(\ref{E:theorem2}), (\ref{E:positive-root}), or even
(\ref{E:regularity}). Instead we must back-track all the way to
(\ref{E:master}).

\subsection{Einstein universe}

The Einstein universe corresponds to
\begin{equation} 
z = -{1\over R^2}, 
\end{equation}
with either sign for the root.  A brief computation leads to
\begin{equation}
\rho={12\over8\pi G_\Newton\;  R^2}; \qquad p= - {4 \over8 \pi G_\Newton\;  R^2},
\end{equation}
and the isotropic form of the Einstein metric:
\begin{equation}
\d s^2 = 
-  \d t^2 
+
{ \d r^2 + r^2 (\d\theta^2 + \sin^2\theta \; \d\phi^2 )
\over
(1+ r^2/R^2)^2}.
\end{equation}
Note that if the density is positive the pressure is negative, and
vice versa --- the Einstein universe does not satisfy the regularity
requirements for a ``normal'' fluid sphere.

\subsection{De~Sitter}

The De Sitter geometry corresponds to
\begin{equation} 
z(r) = {1\over R^2} {3R^2+r^2\over R^2+3r^2}.
\end{equation}
Choose the positive sign for the root.  A brief computation yields
\begin{equation}
\rho=-{12 \over8\pi G_\Newton\; R^2}; \qquad p=  {12 \over8 \pi G_\Newton\; R^2},
\end{equation}
and the isotropic form of the De~Sitter metric:
\begin{equation}
\d s^2 = 
-   \left({1+r^2/R^2\over1- r^2/R^2}\right)^2  \d t^2 
+
{ \d r^2 + r^2 (\d\theta^2 + \sin^2\theta \; \d\phi^2 )
\over
(1- r^2/R^2)^2}.
\end{equation}
%

\subsection{anti-De~Sitter}

The anti-De Sitter geometry corresponds to
\begin{equation} 
z(r) = -{1\over R^2} {3R^2-r^2\over R^2-3r^2}.
\end{equation}
Choose the \emph{negative} sign for the root in (\ref{E:regularity}).
A brief computation yields
\begin{equation}
\rho={12 \over8\pi G_\Newton\; R^2}; \qquad p= - {12 \over8 \pi G_\Newton\; R^2},
\end{equation}
and the isotropic form of the anti-De~Sitter metric:
\begin{equation}
\d s^2 = 
-   \left({1-r^2/R^2\over1+ r^2/R^2}\right)^2  \d t^2 
+
{ \d r^2 + r^2 (\d\theta^2 + \sin^2\theta \; \d\phi^2 )
\over
(1+ r^2/R^2)^2}.
\end{equation}
%

\subsection{The general quadratic ansatz}

Suppose we consider the general quadratic ansatz
\begin{equation} 
z(r) = {A_1 + B_1 \;r^2\over C_1 + D_1 \;r^2}.
\end{equation}
We were led to this ansatz by considering the form of $z(r)$ for the
de~Sitter and anti-de~Sitter cases.  By construction, for {\emph{any}}
choice of $A_1$, $B_1$, $C_1$, and $D_1$ we get a real metric
satisfying the perfect fluid constraint. (Due to rescaling invariance
[multiply both numerator and denominator by any fixed constant], only
three of these four coefficients are actually physically meaningful.)
If the central pressure is to be positive then we must have
$z(0)=A_1/B_1>0$. Then we can without loss of generality take
\begin{equation} 
z(r) = A_2^{\;2} \;{1 + B_2 \;r^2\over 1 + C_2 \;r^2}.
\end{equation}
Now in order to justify calling the geometry an ``exact solution'' we
need an explicit formula for the metric. Inserting this quadratic
ansatz into (\ref{E:theorem2}) the integrals can be done in closed
form. Expressed in terms of $A_2$, $B_2$, and $C_2$ the resulting
metric is a rather messy combination of quadratics (in $r$) raised to
various real exponents.

It is much more convenient to introduce new parameters $S$, $R$, and
$n$ and write
\begin{eqnarray}
\label{E:3parameter}
\d s^2 
&=& 
-\left( {1 \pm r^2/S^2 \over 1 \pm r^2/R^2} \right)^{2n/(2n^2-1)} \; \d t^2
\nonumber
\\
&&
+ \left( {1 \pm r^2/S^2 \over 1\pm r^2/R^2} \right)^{(2n^2-2n+1)/(2n^2-1)} 
\;
{ \d r^2 + r^2 (\d\theta^2 + \sin^2\theta \; \d\phi^2 )
\over
(1\pm r^2/S^2)(1\pm r^2/R^2)}.
\end{eqnarray}
This is a perfect fluid solution for arbitrary $S$, $R$, and $n$;
there are additionally two {\emph{independent}} sign choices that can
be made, one associated with each of the parameters $S$ and $R$. For
definiteness of presentation we discuss the case $++$; but it is
trivial to flip the signs as required.

This solution can (after yet another redefinition of parameters) be
seen to be equivalent to the Goldman I solution~\cite{gold-I}, called
Gold--I in the Delgaty--Lake~\cite{lake} classification. The pressure
and density are rational functions of position.  A brief computation
yields
\begin{eqnarray}
p &=& 
{G_{\hat r\hat r}\over8\pi G_\Newton}  = {G_{\hat\theta\hat\theta} \over8\pi G_\Newton} = 
{G_{\hat\phi\hat\phi} \over8\pi G_\Newton} 
\nonumber
\\
&=&
{ 4 \; q_1(r) \over 
8\pi G_\Newton \; (2n^2-1)^2\; R^4 S^4 \; g_{rr}(r) \; (1+ r^2/S^2)^2\;(1+ r^2/R^2)^2}.
\end{eqnarray}
This verifies that it is
a perfect fluid solution. Here $q_1(r)$ is the quartic
\begin{eqnarray}
q_1(r) &=&  
[(2n^2-1)S^2R^2(R^2-2n^2 S^2)] 
\nonumber\\
&&
+[n^2(R^2-S^2)^2-2S^2R^2(2n^2-1)^2]r^2
\nonumber\\
&&
-[(2n^2-1)(2n^2R^2- S^2)]r^4.
\end{eqnarray}
Similarly
\begin{equation}
\rho = {G_{\hat t\hat t}  \over8\pi G_\Newton}=
{ 4 \; q_2(r) \over 
8\pi G_\Newton \; (2n^2-1)^2\; R^4 S^4 \; g_{rr}(r) \; (1+ r^2/S^2)^2\;(1+ r^2/R^2)^2}
\end{equation}
where $q_2(r)$ is the quartic
\begin{eqnarray}
q_2(r) &=& 
[3(2n^2-1)S^2R^2([n-1]R^2 +n[2n-1]S^2)]  
\nonumber\\
&&
+[n(n-1)(2n-1)(R^2-S^2)^2 +6(2n^2-1)^2S^2R^2]r^2
\nonumber\\
&&
+[3(2n^2-1)(n[2n-1]R^2+ [n-1]S^2)]r^4.
\end{eqnarray}
A somewhat simpler quartic is obtained if we consider
\begin{equation}
\rho+3p = {G_{\hat t\hat t} + 3 G_{\hat r\hat r} \over 8\pi G_\Newton} =
{ 4n(R^2-S^2)\; q_3(r)
\over 
8\pi G_\Newton\;  (2n^2-1)^2 \;  R^4 S^4 \; g_{rr}(r) \;(1+ r^2/S^2)^2\;(1+ r^2/R^2)^2},
\end{equation}
since then
\begin{equation}
q_3(r) =  
[3 (2n^2-1)R^2S^2] + [(2n^2+1) (R^2-S^2) ]r^2 - [3(2n^2-1)]  r^4.
\end{equation}

In particular, at the center of the star
\begin{equation}
p_c = {1\over8\pi G_\Newton} \;
{R^2-2n^2S^2\over (2n^2-1) S^2 R^2},
\end{equation}
while
\begin{equation}
\rho_c = {1\over8\pi G_\Newton} \;
{12[(n-1)R^2 + n(2n-1)S^2]  
\over (2n^2-1) S^2 R^2},
\end{equation}
and
\begin{equation}
\rho_c +3p_c = {1\over8\pi G_\Newton} \;
{12n(R^2-S^2)  
\over (2n^2-1) \; S^2 R^2}.
\end{equation}
From this we can use the positivity of central pressure and density to
constrain the parameters.

We can find the surface of the star by locating the first zero $p(r)$
[or $q_1(r)$], and then use the Hernandez--Misner formula to deduce the
total mass. While a closed-form exact algebraic mass formula exists,
it is too unwieldy to be reproduced here.

To summarize the situation so far: The generating function technique
developed in this note has helped us in several ways. It led us to
consider the quadratic ansatz, realise it was explicitly integrable,
and find a simple form for the metric. We shall now show that this
quadratic ansatz (the Gold--I solution) is also equivalent to the G--G
solution, and that it furthermore contains many interesting special
cases: interior Schwarzschild, Stewart, Kuch5 XIII, de Sitter, anti-de
Sitter, and Einstein among them.

\subsection{Glass--Goldman: G--G}

The G--G~\cite{glass} geometry (Glass--Goldman), in the form reported
by Delgaty--Lake~\cite{lake}, is
\begin{eqnarray}
\d s^2 &=& 
- 
\left(
{2r^2-B^2D-1-C
\over
2r^2-B^2D-1+C}
\right)^{2B/C} \;
 \d t^2 
\nonumber
\\
&&
+
{1\over B^2(2+D)-(B^2D+1)r^2+r^4}
\left(
{2r^2-B^2D-1-C
\over
2r^2-B^2D-1+C}
\right)^{-(B^2D+2B-1)/C}
\nonumber
\\
&&
\qquad\qquad\times 
\left[ \d r^2 + r^2 (\d\theta^2 + \sin^2\theta \; \d\phi^2 ) \right],
\end{eqnarray}
where 
\begin{equation}
C= \sqrt{(B^2D-1)^2-8B^2}
\end{equation}
Thus at first glance it appears to be a two-parameter solution to the
perfect fluid field equations.  There is a subtlety here in the fact
that G--G have implicitly chosen their $r$ coordinate to be
dimensionless, effectively hiding a dimensional parameter in their
conventions, that is
\begin{equation}
r_{\mathrm{this~paper}} = \kappa_0 \; r_{\mathrm{GG}}
\end{equation}
with $\kappa_0$ some arbitrary but fixed distance scale. Then
translating the $B$, $D$, $\kappa_0$ variables to our notation
\begin{eqnarray}
S^2&=&  -\kappa_0^2 \; (1+B^2D+C)/2;
\\
R^2&=&  -\kappa_0^2 \; (1+B^2D-C)/2;
\\
n(R,S) &=& \sqrt{ \half {R^2+\kappa_0^2\over S^2+\kappa_0^2}};
\qquad
\hbox{or}
\qquad
\sqrt{ \half {S^2+\kappa_0^2\over R^2+\kappa_0^2}};
\end{eqnarray}
depending on one's choice for the sign of the square root in the
definition of $C$.  That is, despite appearances the G--G solution is
equivalent to the Gold--I solution and is equivalent to our general
quadratic ansatz.

From the point of view of (\ref{E:3parameter}) this is a reflection of
the fact that that solution is scale-\emph{covariant} under
$r\to\lambda r$, $S\to\lambda S$, $R\to\lambda R$.

\subsection{Schwarzschild Interior Geometry}

The Schwarzschild Interior geometry is a special case of the quadratic
ansatz. It corresponds to taking both sign-choices positive $++$, and
setting $n=1$.  It is now easy to check that the metric is
\begin{equation}
\d s^2 = 
- 
{(1+r^2/S^2)^2
\over
(1+ r^2/R^2)^2} \;
 \d t^2 
+
{ \d r^2 + r^2 (\d\theta^2 + \sin^2\theta \; \d\phi^2 )
\over
(1+ r^2/R^2)^2}.
\end{equation}

A brief computation yields
\begin{equation}
\rho={1\over8\pi G_\Newton} \; {12\over R^2}; 
\qquad 
p= {1\over8\pi G_\Newton} \;
{4\over S^2 R^4}  
{R^2(R^2-2S^2)-(2R^2-S^2)r^2
\over1+r^2/S^2 }.
\end{equation}
The central pressure is
\begin{equation}
p_c =  {4 \over8 \pi G_\Newton}  \; {R^2-2S^2\over R^2S^2 }.
\end{equation}
The stellar surface is located at
\begin{equation}
r_{\mathrm{surface}} = R \; \sqrt{R^2-2S^2\over2R^2-S^2}.
\end{equation}
The total mass is
\begin{equation}
m = {2\over27} \; R \left[{(2R^2-S^2)^{3/2} \; (R^2-2S^2)^{3/2}\over (R^2-S^2)^3}\right].
\end{equation}
We mention that the generating function is
\begin{equation} 
z(r) = {R^2-2S^2-r^2\over R^2 S^2 +(2R^2-S^2) r^2}.
\end{equation}

Note that $S\to 0$ is a singular limit of the interior Schwarzschild
geometry where the central pressure goes to infinity; the central core
of the stellar model is on the verge of becoming a black hole.  On the
other hand, as $S^2\to R^2/2$ from below the stellar surface moves
inward and the star vanishes.

If we drive $S$ out of this ``regular'' range and in particular force
$S^2\to\infty$ then one obtains the Einstein universe (see above).
Finally, $S^2\to -R_0^2$ and $R^2\to +R_0^2$ (that is, the $-+$ sign
choice) corresponds to the anti-De~Sitter universe with scale factor
$R_0$ (see above); while $S^2\to +R_0^2$, $R^2\to -R_0^2$ (that is
$+-$) corresponds to the De~Sitter universe with scale factor $R_0$.

\subsection{Stewart}

To obtain Stewart's geometry~\cite{stewart} we choose the signs to be
$--$ and pick $n=-1$; it is also convenient (but not mandatory) to
interchange the roles of $R$ and $S$.  When written in this form we
can see that it is very closely related to the interior Schwarzschild
solution.  It is now easy to check that the metric is
\begin{equation}
\d s^2 = 
- 
{(1-r^2/S^2)^2
\over
(1- r^2/R^2)^2} \;
 \d t^2 
+
{(1-r^2/R^2)^4
\over
(1- r^2/S^2)^6} 
\; \left[ \d r^2 + r^2 (\d\theta^2 + \sin^2\theta \; \d\phi^2 ) \right].
\end{equation}

A brief computation yields
\begin{equation}
\rho={1\over8\pi G_\Newton}\;
{12R^8(S^2-r^2)^4[S^2(2S^2-3R^2)-(2R^2-3S^2)r^2]
\over
S^{12} (R^2-r^2)^5};
\end{equation}
and
\begin{equation}
p= {1\over8\pi G_\Newton} \;
{4R^8(S^2-r^2)^4[S^2(2R^2-S^2)+(R^2-2S^2)r^2]
\over
S^{12} (R^2-r^2)^5}.
\end{equation}
The central pressure is
\begin{equation}
p_c =  {4 \over8 \pi G_\Newton} \;  {2R^2-S^2\over R^2S^2 }.
\end{equation}
Which implies $R^2>S^2/2$. In contrast, the central density is
\begin{equation}
\rho_c =  {12 \over8 \pi G_\Newton} \;  {2S^2-3R^2\over R^2S^2 },
\end{equation}
which implies $S^2>3R^2/2$. Combined this provides a rather tight
constraint
\begin{equation}
\half S^2 < R^2 < {2\over 3} S^2.
\end{equation}
The stellar surface is located at
\begin{equation}
r_{\mathrm{surface}} = S \sqrt{2R^2-S^2\over2S^2-R^2}.
\end{equation}
The total mass is
\begin{equation}
m = {2\over27} \; S \left[{(2S^2-R^2)^{3/2} \; (2R^2-S^2)^{3/2}\over R^4 (R^2-S^2)}\right].
\end{equation}
We mention that the generating function is
\begin{equation} 
z(r) = {2R^2-S^2-r^2\over R^2 S^2 +(R^2-2S^2) r^2};
\end{equation}
%

\subsection{Kuchowicz: Kuch5 XIII}

To obtain the Kuch5 XIII geometry~\cite{kuch5XIII} we simply let
$R\to\infty$; it is then convenient (but not mandatory) to re-label
$S$ as $R$.  It is now easy to check that the metric is
\begin{equation}
\d s^2 = 
- 
(1+ r^2/R^2)^{2n/(2n^2-1)}
 \d t^2 
+
(1+ r^2/R^2)^{(2-2n)/(2n^2-1)}
\; \left[ \d r^2 + r^2 (\d\theta^2 + \sin^2\theta \; \d\phi^2 ) \right].
\end{equation}
A brief computation yields
\begin{equation}
\rho={1\over8\pi G_\Newton} \;
{4(n-1)[3(2n^2-1)+(2n^2-n)r^2/R^2] \; (1+r^2/R^2)^{-2n(2n-1)/(2n^2-1)}
\over
(2n^2-1)^2\;R^2}.
\end{equation}
and
\begin{equation}
p= {1\over8\pi G_\Newton} \;
{4[(2n^2-1)+n^2 r^2/R^2] \; (1+r^2/R^2)^{-2n(2n-1)/(2n^2-1)}
\over
(2n^2-1)^2\;R^2}.
\end{equation}

The central pressure is
\begin{equation}
p_c =  {4 \over8 \pi G_\Newton}  \;  {1\over R^2(2n^2-1) }.
\end{equation}
Which implies $n^2>1/2$. In contrast, the central density is
\begin{equation}
\rho_c =  {12 \over8 \pi G_\Newton} \;  {n-1\over R^2\; (2n^2-1) },
\end{equation}
which further implies $n>1$.  Because of these constraints there is no
stellar surface; pressure remains positive for all values of $r$ and
the solution is actually cosmological. (This ultimately can be traced
back to the fact that $B_2=0$; which means we are dealing with a
singular solution of our general three-parameter result.)

We mention:
\begin{equation} 
z(r) = {1\over R^2(2n^2-1)+2n^2 r^2}.
\end{equation}
Also, it is formally possible to replace $R^2\to-R^2$ at the cost of
reversing the positivity conditions ($2n^2<1$; $n<1$).

\section{Discussion and Conclusions}
\setcounter{equation}{0}

We have explicitly characterized the spacetime metrics corresponding
to the class of all static spherically symmetric perfect fluid
geometries in a relatively straightforward manner. This observation is
useful whenever there is some uncertainty regarding the actual
equation of state one wishes to use. The first theorem we presented is
applicable to all static spherically symmetric perfect fluid
geometries without further restriction, while the second theorem
encodes the most important of the regularity conditions that are
relevant to an isolated static fluid sphere (such as a star).

Though the formulae we present do involve an integration, it is
particularly noteworthy that in our representation the metric is
explicit. Furthermore it is easy to keep the metric real, and
particularly easy to find the surface of the ``star''.  Some
(but not all) of the standard regularity conditions are easy to
enforce, and can be interpreted as extra restrictions on the class of
``generating functions'' $z(r)$.

Throughout this article we chose to work in isotropic coordinates,
because we found them to be the most useful. (See Glass and Goldman
for an earlier, and rather different, use of the ideas of isotropic
coordinates and generating functions~\cite{glass}.) The use of
isotropic coordinates is not a matter of deep principle and we do not
rule out the possibility that there may still be other (possibly even
simpler) representations in other coordinate systems. For instance,
the recent work of Fodor~\cite{fodor} in Schwarzschild coordinates is
particularly intriguing.

In closing we reiterate that while a tremendous amount is already
known concerning static spherically symmetric spacetimes (see in
particular~\cite{Exact,lake,skea}) the particular approach adopted in
the present article falls well outside any of the standard schemes.\footnote{%
Among the approaches we have encountered in the literature, the
closest in spirit to the current approach is the Baumgarte--Rendall
analysis of~\cite{Baumgarte}.}

\section*{Acknowledgments}

The research of Matt Visser was supported by the US DOE. Shahinur
Rahman wishes to acknowledge support from Washington University via
the ``Mr. and Mrs. Nicolas M. Georgitsis Scholarship''. The authors
wish to thank the referees for their comments, and in particular for
drawing our attention to the importance of reference~\cite{Baumgarte}.

The authors also particularly wish to thank Kayll Lake for pointing
out an error in earlier versions of our mass formulas, and for putting
us on the right track.

 

\end{document}